\documentstyle[preprint,revtex]{aps}
\begin{document}
\draft

\begin{title}
{RESPONSE FUNCTIONS FROM \\
 INTEGRAL TRANSFORMS WITH A LORENTZ KERNEL\\}
\end{title}

\vskip 3 truecm

\author{Victor D. Efros}
\begin{instit}
Russian Research Centre "Kurchatov Institute"\\
 Kurchatov Square, 1, 123182 Moscow, Russia
\end{instit}
\author{Winfried Leidemann}
\begin{instit}
European Centre for Theoretical Nuclear Physics and Related Areas\\
Villa Tambosi, I-38050 Villazzano (Trento) Italy
\end{instit}
\centerline{and}
\author{Giuseppina Orlandini}
\begin{instit}
Dipartimento di Fisica, Universit\`a di Trento, I-38050 Povo
(Trento) Italy\\and Istituto Nazionale di Fisica Nucleare,
Gruppo Collegato di Trento
\end{instit}
\vfill\eject

\vskip 5 truecm

\begin{abstract}
We propose to calculate inelastic response functions
from the inversion of their integral
transform with a Lorentz kernel. The transform can be obtained
using bound-state type methods. Thus one does not need to solve the much
more complicated continuum problem with many open channels.
Contrary to other integral transforms considered in the literature,
the inversion leads to results of excellent accuracy and stability.
This is explicitly shown for the longitudinal deuteron response.
\end{abstract}

\vfill \eject
The calculation of nuclear response functions from realistic nuclear dynamics
is of great importance. It would allow checking
the generally accepted form of the electromagnetic operators and help
clarifying
the issue of modification of nucleon properties inside the nucleus. However,
in the framework of the conventional approach that deals with many-body
continuum wave functions, such calculations are out of reach at present.
Only very recently 3N continuum-state solutions with
realistic N-N forces have been applied to the exclusive electron induced
$^3$He break up \cite{gloeckle}. Similar realistic calculations for the
inclusive reaction are far more
laborious since one should sum a great number of break up contributions.
Such results were obtained in a single calculation \cite{MeT}
with a simplified $S$-wave central N-N force only.

The possibility of obtaining response
functions of few-body systems from the inversion of
integral transforms has been investigated in Refs.
\cite{efros85,carlsonschiavilla92,stielt193}. The essential point is that
these transforms can be evaluated  without knowing the continuum wave
functions.
This point is not a technical one but on the contrary of great importance as
it leads to an enormous simplification of
the calculation. In particular, it allows obtaining realistic
response functions for 4-body systems and it considerably simplifies 3-body
calculations. Stieltjes and
Laplace transforms have already been considered, however, the inversion of
these
transforms is very cumbersome. Particularly in the quasielastic peak region
one encounters great problems, since responses with rather different peak
shapes
can lead to almost equal Stieltjes and Laplace transforms
\cite{stielt193,LeO93}. Thus it is very difficult
to get a unique result for the response from their inversion.

In this letter we propose to evaluate the response function $R(\omega)$ at
constant momentum transfer from inverting their integral transform  with a
Lorentz kernel
\begin{equation}
\Phi(\sigma_R,\sigma_I)= \int_{\omega_{min}}^\infty d\omega
{R(\omega)\over \sigma_I^2+(\omega - \sigma_R)^2} \,, \,\,\,\,\,
\sigma_R,\,\sigma_I > 0 \,.
\end{equation}
(Contributions to $\Phi$ due to transitions to discrete levels are omitted,
they can be separated out as done in Ref. \cite{stielt193}).
The g\rangle\,.
\end{equation}

The last expression of Eq. (2) shows that, contrary to continuum-state wave
functions, $\Psi_0$ {\em vanishes} at large
distances like a bound-state.
This defines the boundary condition to Eq. (3), which is much
simpler to solve than the continuum equation with many open channels.

In the following we show that the proposed method actually works. To this
end we
(a) construct the transform for the longitudinal response of the deuteron
using Eqs. (3) and (2), (b) invert the obtained transform  and (c) compare the
resulting $R(\omega)$ with the one of a conventional calculation with
explicit use of continuum wave functions.
In strict analogy to Ref. \cite{stielt193}
and following the notations there, one has
\begin{equation}
\Phi(\sigma_R,\sigma_I) = {1\over 3}\sum_L\sum_{jl}(2 j + 1) \int_0^\infty dr
\,|\phi_{L,jl}(\sigma_R,\sigma_I,r)|^2\,.
\end{equation}
The complex $\phi_{L,jl}$ is obtained from the same radial equations\
as Eqs. (27) and (28) of Ref. \cite{st to 20\%
for $\sigma_I = 5$ MeV. The increase of the error for decreasing $\sigma_I$
and increasing $\sigma_R$ is understood comparing our approximate solution
with the true asymptotic
 $\phi_{L,jl}$, which behaves
  - up to small corrections - as
${\rm exp}[(2M\sigma_R/\hbar^2)^{1/2}\{[-i -\sigma_I/(2\sigma_R)]r\}].$ At
the same $r_{asy}$ value  a better procedure would be matching
 $\phi_{L,jl}$ with this  asymptotic solution. But we renounce to improve
 the accuracy just to show the nice stability of our
 method. In fact, we will show that using the rather approximate
 $\Phi(\sigma_R,\sigma_I)$ obtained as an input to our integral equation does
 not automatically lead to inversion problems like for Stieltjes and
 Laplace transforms.

The response $R(\omega)$ is obtained from the calculated
$\Phi(\sigma_R,\sigma_I)$ through the
inversion of Eq. (1). To this aim we expand $R(\omega)$ in a set of
functions $\chi_n(\omega)$:

\begin{equation}
R(\omega) = \sum_{n=1}^{N} a_n\, \chi_n(\omega)\,\,\,,\,\,\,\,
\chi_n(\omega) = {(\omega-E_b)}^{n- {1 \over 2}}\exp({-(\omega-E_b)
\over E_1})\,\,,
\end{equation}
where $E_b$ is the deuteron binding energy and $E_1$ is
a free parameter. Eq. (5)
corresponds to the following expansion of the
transform
\begin{equation}
\Phi_{trial}(\sigma_R,\sigma_I) = \sum_{n=1}^N a_n\, \xi_n(\sigma_R,
\sigma_I)\,\,,\,\,\,\,\,
\xi_n(\sigma_R,\sigma_I)=\int_{\omega_{min}}^\infty {\chi_n(\omega)
\over\sigma_I^2+(\omega - \sigma_R)^2}\,d\omega\,\,.
\end{equation}
For a fixed value of $\sigma_I$ we calculate $\Phi(\sigma_R,\sigma_I)$
in  $I\gg N$ points $\sigma_R(i)$,
the $a_n$ and $E_1$ are then determined from a best fit to
$\Phi(\sigma_R(i),\sigma_I)$. Since $R(\omega)$ has to be
positive we discard all solutions with $R(\omega) < 0$. Fig. 1b shows
the result of the inversion. The solutions are extremely stable: given a
sufficiently large $N$ $(N \geq 10)$ we find almost identical results for
any $E_1$. The only visible differences are present at low energies, where one
has rather strong oscillations. Apart from this region one finds an
excellent agreement with the exact $R(\omega)$. One also notes that the
low-energy oscillations are smaller for the smaller $\sigma_I$,
which had to be expected because of the better resolution.

We improve the low-energy results as follows.
As Eq. (4) shows,
$\Phi$ is obtained as a sum of $L$-multipole transitions to partial waves
with total spin $j$ and orbital
momentum $l$. Analyzing their $\sigma_R$ dependence separately we find that
almost all of them  have only one structure due to the quasielastic peak.
Only the $L=0,$ 2 transitions to the state with $j=1$ and $l=0,$ 2 have
strength at low $\sigma_R$.
Fig. 2a shows the results of their inversion.
Now the agreement with the exact results is reasonably good also
at low energy. For all the remaining contributions we invert the integral
transform of their sum. Adding to this the two separate contributions of Fig.
2a
leads to an excellent final result as shown in Fig. 2b.
The improvement may be understood as follows. We  have an integral equation
of the first kind like e.g. the Fourier transform, whose solution is
unstable against adding high-frequency components to it. It means that only
a limited number $N$
   of
basis functions may be retained in Eq. (6) and the higher is the accuracy
of $\Phi$ the larger $N$ are permissible. It may be hard to represent all the
structures of $R$ with a small $N$, but since the separate contributions
have less structure they are easier described by a small $N$. We also
mention that we come to almost identical results using quite another set of
basis functions (set $\chi^{(2)}$ of Ref. \cite{stielt193}) in the expansion
(5).

We also want to estimate accessible $\sigma_I$ values. This is
important for 3- and 4-body calculations. Any $\sigma_I$ would be acceptable
if the dynamical input $\Phi$ to Eq. (1) were calculated exactly.
It is expedient to use $\sigma_I$
values which are less than the widths of typical structures in the response.
For few-body responses such
values are about 10 MeV as the above calculation confirms.
On the other hand, for $\sigma_I \rightarrow 0$  the solution  of Eq. (3)
becomes very long ranged with all open channel asymptotics being present at
large distances. It is hard to get such a solution with a good accuracy
using bound-state-type methods. So the question arises whether the 10 MeV are
sufficiently large. We discuss this
point for the method of hyperspherical harmonics. In order to treat properly
interparticle distances $R$ generalized momenta $K$
up to
$K_{max}\simeq R/R_0$ are required ($R_0$: size of a subsystem present at
large distances). Considering this and using the
asymptotics with the exponent given above, but substituting the
hyperspherical $\rho \simeq R$ for $r$, one estimates $K_{max}\simeq
(\sigma_RE_{cl})^{1/2}/\sigma_I$ ($E_{cl}$: binding energy of an asymptotic
subsystem). The  $K_{max}$ values required for typical  $\sigma_R < 200$ MeV
and $\sigma_I\simeq 10$ MeV  are quite acceptable.

Before concluding we discuss an important point in view
of an application of the present method to more complex nuclei, i.e. a
variational principle for solving Eq. (3).
Let us write it in the short
form $\hat{L} \Psi_0 = Q$ and consider the functional
\begin{equation}
I[\psi ] = <\psi\mid L^{\dagger}L\mid \psi>-<\psi\mid L^{\dagger}Q>-
<Q\mid L\psi>, \label{eq:fun}
\end{equation}
where $\psi$ are localized functions. For its increment
$\Delta I = I[\psi]-I[\Psi_0]$ one has
\begin{equation}
\Delta I = <\delta \psi\mid L^{\dagger}L\mid \delta \psi> \label{eq:var}
\end{equation}
with $\delta \psi =\psi -\Psi_0$. Eq. (\ref{eq:var}) shows
that $\Delta I > 0$  for any $\psi$ different from $\Psi_0$.
So the functional from Eq. (\ref{eq:fun}) is minimal just on the
solution $\Psi_0$ of our dynamical equation (3). Thus it can  be
replaced by the requirement
\begin{equation}
I[\psi ] = \mbox {min}\,.
\end{equation}
This allows applying, e.g., the  variae{
(a) Relative error of $\Phi(\sigma_R,\sigma_I)$ for $\sigma_I = 5$, 10 MeV
(exact $\Phi$ from Eq. (1) with the known $R(\omega)$);
(b) the deuteron longitudinal response as a function of $E_{np}=\omega-E_b$ at
$q_{c.m.}^2=5$ fm$^{-2}$. Solid curve: result from conventional
calculation, dashed and dotted
curves: results of our method with $\sigma_I = 5$, 10 MeV, respectively (N=14,
see Eqs. (5,6)).
\label{Fig.1}}
\figure{
(a) C0 and C2 multipole contributions
to $R(\omega)$ for the transition
to the ${^3S_1-^3D_1}$ partial wave. Solid
curves: results from conventional calculation (shown up
to 100 MeV only), dashed curves:  results of our method
with $\sigma_I = 5$ MeV (N=10, see Eqs. (5,6));
(b) total $R(\omega)$, notations as in (a). Note that the result with our
method
is the sum of three separate contributions (see text).
\label{Fig.2}}

\end{document}